\pdfoutput=1

\documentclass[12pt]{article}
     
\usepackage{epsfig}
\usepackage{hyperref}
\usepackage{graphicx}
\usepackage{amsmath}
\usepackage[numbers]{natbib}

\begin{document}

\title{Limits on hadron spectrum \\from bulk medium properties%
\thanks{Supported by the Polish National Science Centre grants 
DEC-2015/19/B/ST2/00937 and DEC-2012/06/A/ST2/00390. }}

\author{Wojciech Broniowski$^{1,2}$\\
~\\
$^1$ Institute of Physics, Jan Kochanowski University \\ 25-406 Kielce, Poland \\
$^2$ The H. Niewodnicza\'nski Institute of Nuclear Physics \\ Polish Academy of Sciences, 31-342 Cracow, Poland}

\date{Presented at Mini-Workshop Bled 2016: \\ QUARKS, HADRONS, MATTER, Bled (Slovenia),  July 3-10, 2016 }
                    
\maketitle

\begin{abstract}
We bring up the fact that the bulk thermal properties of the hadron gas, as measured on the lattice, preclude a very fast rising of the
number of resonance states in the QCD spectrum, as assumed by the Hagedorn hypothesis, unless a substantial repulsion 
between hadronic resonances is present. If the Hagedorn growth continued 
above masses $\sim 1.8$~GeV, then the thermodynamic functions would noticeably depart from the measured lattice values 
at temperatures above $140$~MeV, just below the transition temperature to quark-gluon plasma.
\end{abstract}

\bigskip 
  
In this talk we point out the sensitivity of thermal bulk medium properties (energy density, entropy, sound velocity...)
to the spectrum of the hadron resonance gas. In particular, we explore the effects of the high-lying part of the 
spectrum, above $\sim 1.8$~GeV, where it is poorly known, on the thermal properties still below the 
cross-over transition to the quark-gluon plasma phase. Such investigations were carried out in the past 
by various authors, see~\cite{Karsch:2003vd,Huovinen:2009yb,Andronic:2012ut,Arriola:2014bfa,Lo:2015cca} and references therein, where 
the reader may find more details and results.

\begin{figure}
\begin{center}
\includegraphics[width=0.7 \textwidth]{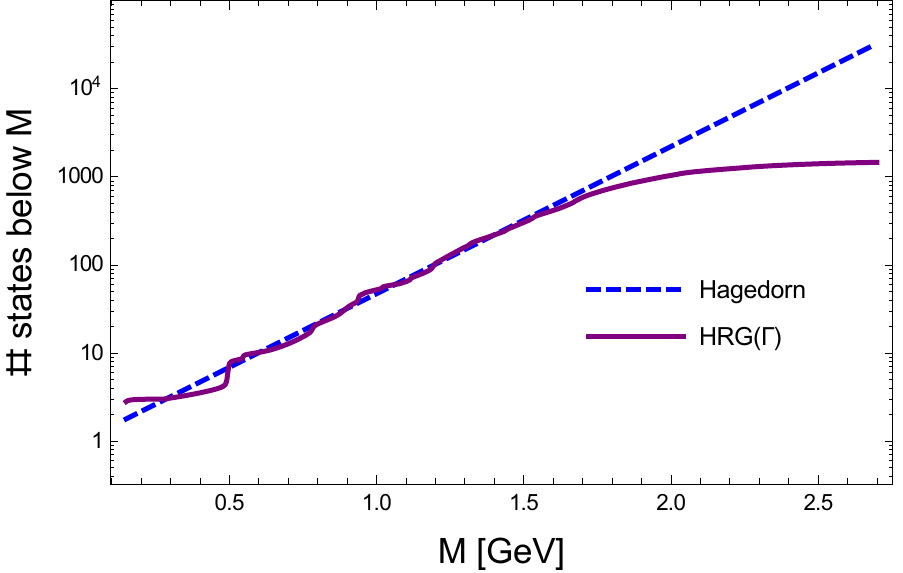} 
\end{center}
\vspace{-3mm}
\caption{Solid line, labeled HRG($\Gamma$): Number of QCD states (mesons, baryons, and antibaryons combined) with mass below $M$.
All stable particles and resonances from the Particle Data Group tables~\cite{Olive:2016xmw} are included and their Breit-Wigner width is 
taken into account. 
Dashed line: the fit with the Hagedorn formula for the density of states, $\rho(m)=A \exp(m/T_H)$, with $T_H=260$~MeV.
\label{fig:hag}}
\end{figure}

The presently established QCD spectrum reaches about 2~GeV, and it is a priori not clear what happens above. Does the growth 
continue, or is saturated? As is evident from Fig.~\ref{fig:hag}, the Hagedorn hypothesis~\cite{Hagedorn:1965st} works very well up to about 1.8~GeV~\cite{Broniowski:2004yh}.
In the following, we explore two models: 1)~hadron resonance gas with the Breit-Wigner width, HRG($\Gamma$), which takes into account all states listed 
in the  Particle Data Group tables~\cite{Olive:2016xmw} with mass below 1.8~GeV, and 2)~this model amended with the states above 1.8~GeV, modeled with the 
Hagedorn formula fitted to the spectrum at lower masses (see Fig.~\ref{fig:hag}). In short, model~1) includes the up-to-now established resonances, and model~2)
extends them according to the Hagedorn hypothesis.

\begin{figure}
\begin{center}
\includegraphics[width=0.7 \textwidth]{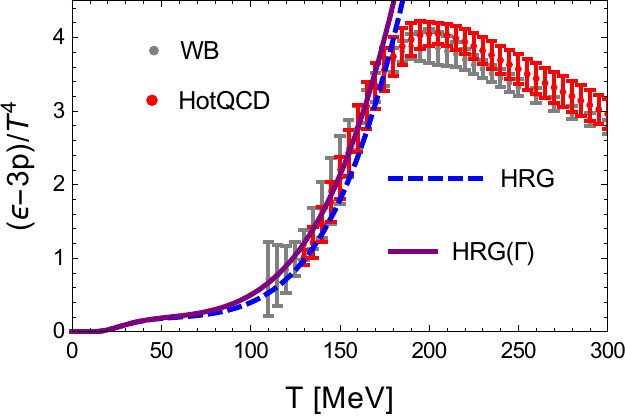} 
\end{center}
\vspace{-3mm}
\caption{The QCD trace anomaly (divided by $T^4$) plotted as a function of temperature $T$. The inclusion of 
width of resonances to the hadron resonance gas improves the agreement with the lattice data from the 
Wuppertal-Budapest (WB)~\cite{Borsanyi:2013bia} and Hot QCD~\cite{Bazavov:2014pvz} collaborations. \label{fig:tr}}
\end{figure}

First, we recall 
the fact that the inclusion of widths of resonances~\cite{Arriola:2012vk}, 
as listed in the Particle Data Group tables, affects the results noticeably and in fact improves them.  
This is shown in Fig.~\ref{fig:tr}, where the hadron resonance gas calculation for the QCD trace anomaly, $\epsilon -3p$ divided by $T^4$. Here
$\epsilon$ stands for the energy density, $p$ for the pressure, and $T$ for the temperature. In the calculation, 
the hadrons are treated as components of an ideal gas of fermions and bosons. We note that 
the overall agreement with of the hadron resonance gas model HRG($\Gamma$) with the lattice 
measurement is remarkable. 

The virial expansion of Kamerlingh Onnes yields $p/T=\rho + B_2(T)\rho^2+  B_3(T)\rho^3+\dots$.
Correspondingly, for the partition function of a thermodynamic system including the $1\to 1$, $2\to 2$, etc., processes one has
\begin{eqnarray} 
\ln Z= \ln Z^{\rm (1)}+\ln Z^{\rm (2)}+\dots
\end{eqnarray}
The non-interacting term
\begin{eqnarray}  
\ln Z^{\rm (1)} = \sum_k \ln Z^{\rm stable}_{k}=  \sum_k  f_{k} V \int  \frac{d^{3}p}{(2\pi)^{3}}\ln\left[  1\pm e^{-E_p/T}\right]^{\pm 1}
\end{eqnarray}
includes the sum over all stable particles, 
whereas the second-order virial term involves the sum over pairs of stable particles denoted as $K$,
\begin{eqnarray}
\ln Z^{\rm (2)} = \sum_K f_{K} V \int_{0}^{\infty} \frac{d \delta_K(M)}{\pi dM}\, dM\int \frac{d^{3}P}{(2\pi )^{3}}\ln\left[  1\pm e^{-E_P/T}\right]^{\pm 1},
\label{eq:z2}
\end{eqnarray}
where $\delta_K(M)$ stands for the phase shift in the channel $K$. For narrow resonances the correction to the density of two-body states $d \delta_K(M)/(\pi dM)$~\cite{Broniowski:2015oha} 
can be accurately approximated with the Breit-Wigner form, which is a basis of the hadron resonance gas model.

\begin{figure}
\begin{center}
\includegraphics[width=0.7 \textwidth]{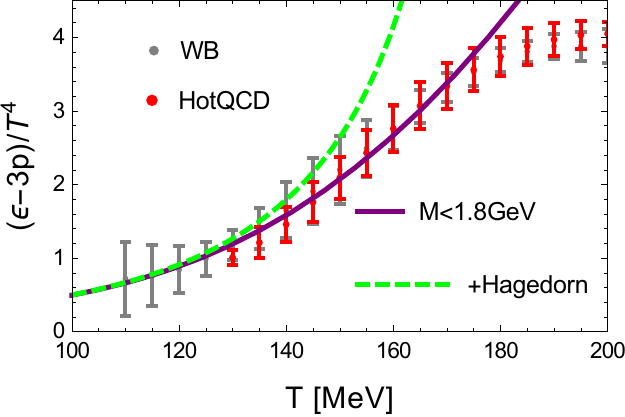} 
\end{center}
\vspace{-3mm}
\caption{Same as in Fig.~\ref{fig:tr} but with the lines denoting the hadron resonance gas model, HRG($\Gamma$), up to $M=1.8$~GeV, and 
this model amended with the Hagedorn spectrum above  $M=1.8$~GeV. \label{fig:tr18}}
\end{figure}

\begin{figure}
\begin{center}
\includegraphics[width=0.7 \textwidth]{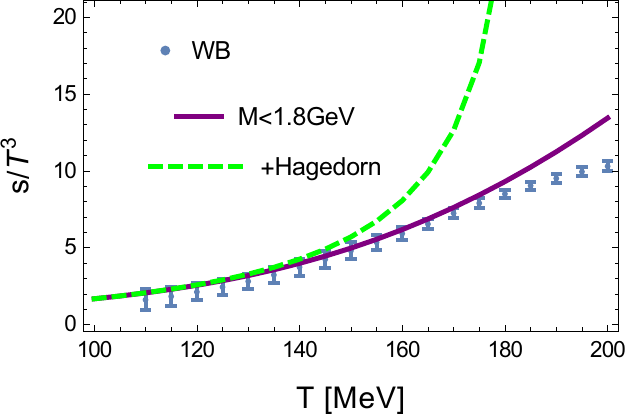} 
\end{center}
\vspace{-3mm}
\caption{Same as in Fig.~\ref{fig:tr18} but for the entropy density divided by $T^3$. \label{fig:s18}}
\end{figure}

\begin{figure}
\begin{center}
\includegraphics[width=0.7 \textwidth]{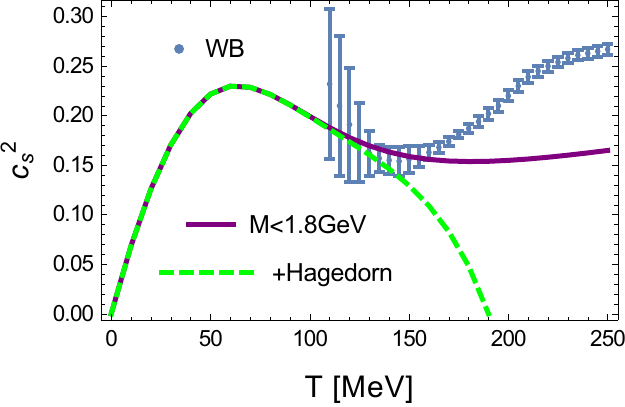} 
\end{center}
\vspace{-3mm}
\caption{Same as in Fig.~\ref{fig:tr18} but for the square of the sound velocity. \label{fig:cs18}}
\end{figure}

In Fig.~\ref{fig:tr18} we show the result of extending the Hagedorn hypothesis above the present experimental limit on the QCD spectrum. We note 
that the inclusion of extra (non-interacting) states above $M=1.8$~GeV has a quite dramatic effect on the trace anomaly 
$\theta^\mu_\mu=\epsilon-3p$, placing it way above the lattice 
data at $T>140$~MeV (the model calculation is credible below $T\simeq 170$~MeV, where a cross-over to the quark-gluon plasma occurs).
A similar conclusion is drawn for other thermodynamic quantities, such as  the entropy (cf. Fig.~\ref{fig:s18}) or the sound velocity (cf. Fig.~\ref{fig:cs18}).

Therefore, if the hadron resonances were non-interacting, there would be no room for extra states above 1.8~GeV in the QCD spectrum.
This conclusion may be affected by repulsion between the states (e.g., the excluded volume corrections), which decreases the contribution to the 
partition function. The issue is discussed 
quantitatively in~\cite{Andronic:2012ut,Arriola:2014bfa}, where a reduction of contributions to the thermodynamic quantities is assessed. The excluded volume
reduces the contribution of resonances, and this makes them possible to appear in the spectrum in an ``innocuous'' way. The effect is 
explicit in Eq.~(\ref{eq:z2}), as repulsion leads to a decrease of the phase shift with $M$, or a negative correction to the density of states $d \delta_K(M)/(\pi dM)$.

\begin{figure}
\begin{center}
\includegraphics[width=0.7 \textwidth]{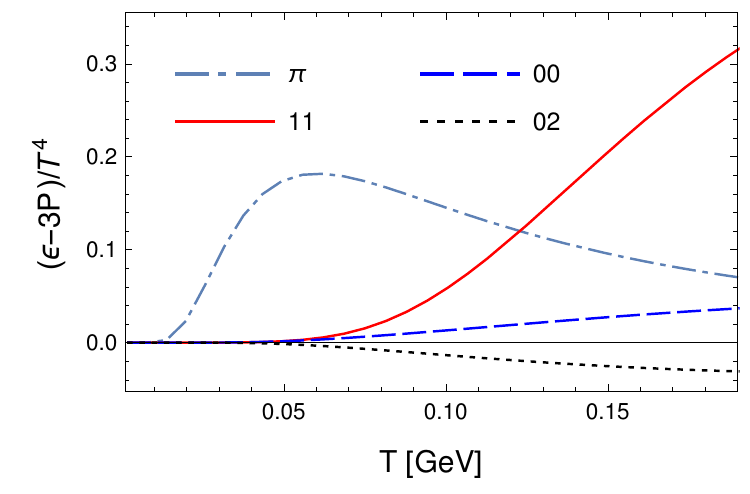} 
\end{center}
\vspace{-3mm}
\caption{Contributions to the trace anomaly from the pions, $\rho$ mesons, $\sigma$ meson, and the isospin-2 component of the pion-pion 
interaction. We note an almost perfect cancellation of the $\sigma$  and isospin-2 channels. \label{fig:tr2}}
\end{figure}

An important example of such an explicit cancellation occurs in the case of the $\sigma$ meson, whose contribution to one-body observables 
is canceled by the isospin-2 channel~\cite{Broniowski:2015oha}. The case of the trace anomaly is shown in Fig.~\ref{fig:tr2}. Note that the 
phase shift taken into account in this analysis automatically includes the short-distance repulsion in specific channel, hence there is no need 
to model it separately. The cancellation experienced by the $\sigma$ state may occur also for other states with higher mass.  

In conclusion, the thermodynamic quantities offered by the modern lattice QCD calculations 
allow to place limits on the high-lying spectrum on the QCD resonances, but the interactions 
between the states, such as the short-range repulsion, must be properly taken into account, as the 
two effects: increasing the number of states and introducing repulsion works in the opposite way.

\bibliography{therm}

\end{document}